\documentclass[12pt,aps]{report}
\renewcommand{\baselinestretch}{1.5}

\usepackage[pdftex]{graphicx}
\begin{document}
\oddsidemargin = -0.52cm     
\textwidth = 16.96cm         
\topmargin = 0cm            
\textheight = 23.7cm        
\voffset = -0.54cm          
\paperwidth = 21cm          
\paperheight = 29.7cm       
\renewcommand{\baselinestretch}{1.3}    
\parskip=6pt

\vskip60mm
\title{On Two Models of the Light Pulse Delay in a Saturable Absorber }
 \author{V. S. Zapasskii and G. G. Kozlov}

\maketitle
\begin{abstract}
A comparative analysis of two approaches to description of the
light modulation pulse delay in a saturable absorber is presented.
According to the simplest model, the delay of the optical pulse is
a result of distortion of its shape due to absorption
self-modulation in the nonlinear medium. The second model of the
effect, proposed at the beginning of our century, connects the
pulse delay with the so-called "slow light" resulting from the
group velocity reduction under conditions of the coherent
population oscillations. It is shown that all the known
experimental data on the light pulse delay in saturable absorbers
can be comprehensively described in the framework of the simplest
model of saturable absorber and do not require invoking the effect
of coherent population oscillations with spectral hole-burning and
anomalous modifications of the light group velocity. It is
concluded that the effect of group velocity reduction under
conditions of coherent population oscillations has not received so
far any experimental confirmation, and the assertions about real
observation of the "slow light" based on this mechanism are
groundless.
\end{abstract}

\section*{INTRODUCTION}

Saturable absorption is a basic effect of the incoherent nonlinear
optics. In its simplest form, the effect is revealed as bleaching
of the medium with increasing intensity of the light passing
through it.  This effect became known still in the pre-laser epoch
due to experimental works of Kasler's group on optical orientation
of atoms \cite{1}. Its observation under relatively low light
power densities was possible because of low population relaxation
rates in these systems. Later on, after the advent of laser, this
effect has been studied in great detail \cite{2, 3, 4} and found
application in the laser systems of Q-switching and mode-locking
(see, e.g., \cite{5,6}). Nonlinearity of the saturable absorber is
determined not by nonlinearity of polarizability of the medium at
optical frequencies, but rather by the dependence of optical
constants of the medium (optical absorption) on the light
intensity or on the light power density. One can say that the
carrier frequency of the light field and, moreover, even the fact
of its existence (see, e.g. \cite{7}) play no role in this effect.
Under consideration is, in essence, the dynamics of penetration of
the energy flux through a barrier whose transmissivity depends on
this flux.

Mechanism of the effect is usually determined by the light-induced
changes in populations of the states of the system and, for this
reason, the response of the absorber exhibits a certain
sluggishness associated with the sluggishness of the population
relaxation.  As a result, the light pulse transmitted through the
saturable absorber appears to be distorted or, under certain
conditions, shifted in time. The sign of this shift, in the
general case, can be both positive and negative. All the
regularities of interaction of light with a saturable absorber
have been studied in great detail to early 70's \cite{2, 3, 4}
and, later on, the interest to these effects was, to a
considerable extent, lost.

At the beginning of our century, however, the effects of retarded
response of saturable absorber have been rediscovered and
reinterpreted \cite{8, 9}. In the new interpretation, the pulse
delay was associated with a low group velocity of light
propagating in the {\it linear medium} with a highly steep
refractive index dispersion. The high steepness of the dispersion,
in authors' opinion, was provided by the narrow spectral hole
burnt by laser light in the homogeneously broadened absorption
band under conditions of the coherent population oscillations
(CPO) \cite{10}. In this interpretation, small temporal shifts of
the light pulse passed through the medium ('small' in the scale of
the pulse length, but huge in the scale of time needed to the
light to pass through the sample with the velocity $c$) were
converted into giant factors of the group velocity reduction. With
appearance of the novel model of pulse delay in a saturable
absorber, the slow light technique, which had in its arsenal, at
that time, only the method of electromagnetically induced
transparency, was enriched by one more experimental approach,
which allowed one to demonstrate achievements of the same (or even
higher) level under much simpler experimental conditions at room
temperatures on solid-state objects, highly convenient for
practical applications (see, e.g., \cite{11, 12}).

Thus, it occurred so that one and the same phenomenon was
described in the framework of two, fundamentally different and
incompatible models. The discussion aroused on this issue in
literature \cite{13,14,15,16,17,18,19,20,21} has not lead to a
consensus.

In this paper, we present a comparative analysis of the two models
of the light pulse delay in a saturable absorber and show that the
effect of the CPO-based reduction of the light group velocity has
not been reliably observed so far, all the relevant experimental
data on pulse delay perfectly agree with the simplest model of
saturable absorber, and there are no grounds to consider the
CPO-based slow light as really existing.

\section*{ PULSE DELAY AS A RESULT OF ITS SHAPE DISTORTION IN A SATURABLE ABSORBER}

Most essential properties of a saturable absorber can be treated
neglecting the propagation effects (for more detail, see
\cite{14}). For that, it suffices, in the expression describing in
the general form the nonlinear relation between the intensity of
the light transmitted through the absorber ($I_{out}$) and
intensity of the incident light ($I_{in}$)
\begin{equation}
I_{out} = K(I_{in},t)I_{in},
\end{equation}
to take into account, in the simplest form, the aforementioned
sluggishness of the transmission dynamics of the medium

\begin{equation}
\frac{dK}{dt} = \frac{K_{eq}-K}{\tau}
\end{equation}
($K_{eq}$ is the equilibrium value of the transmissivity $K$  for
the current intensity $I_{in}$), and to restrict oneself to the
linear term of expansion of the transmissivity $K_{eq}$ (implying
relatively small variations of the intensity $I_{in}$)

\begin{equation}
K_{eq}(I_{in}) = K_0 + K_1 I_{in}, K_1 I_{in} \ll K_0,
\end{equation}
Relations (1)--(3), in spite of the used simplifications, make it
possible to describe all the main regularities of response of such
a nonlinear medium to intensity variations of the acting light. In
particular, the frequency dependence of the intensity modulation
amplitude $I_\omega$ exhibits a Lorentzian feature (peak or dip,
depending on sign of the constant $K_1$) with a half-width of ~
$\tau^{-1}$, centered at zero frequency (Fig. 1a), while the
corresponding dependence of the intensity modulation phase shift
$\phi_{\omega}$ shows the greatest delay (respectively, positive
or negative) at the frequency $\omega \sim \tau^{-1}$ (Fig. 1b ).
The light pulse, in the general case, appears to be distorted at
the exit of the medium. However, for a smooth pulse of
sufficiently large duration ($> \tau$),  this distortion is
reduced practically to a pure shift, with its sign being also
determined by the sign of the coefficient $K_1$ (Fig. 1c). In more
detail, all these regularities are described in \cite{2, 4, 14}.

These are the dependences that were demonstrated in the first
publications on the "CPO-based slow light" and that are considered
so far as good evidences for the reduction of the light group
velocity in saturable absorbers.

The dependences presented in Fig. 1 describe unavoidable
properties of a saturable absorber that have nothing to do with
anomalous modifications of the light group velocity in the medium.
In particular, the above effects of pulse delay or delay of the
light intensity modulation can be observed in media of ultimately
small thickness in the light of arbitrary spectral composition.
The pulse delay is, in essence, a result of light self-modulation,
 and the pulse shape changes when the light-induced bleaching (or
darkening) of the medium occurs in the process of the pulse
propagation, and absorption of the medium for the leading and
trailing edges of the pulse proves to be different.

It is noteworthy that Eqs. (1)--(3) describe, along with
transformation of temporal behavior of the light transmitted
through a saturable absorber, transformation of its {\it intensity
spectrum}. The saturable absorber, with respect to the transmitted
light, plays the role of the light-controlled optical modulator.
However, in view of Eq. (3), high frequencies of the light
intensity modulation do not participate in controlling the
modulator. For this reason, the main distortions of the intensity
spectrum occur in the range of low frequencies ($\omega <
\tau^{-1})$. In terms of the conventional spectroscopic technique,
the most direct way to make sure that this is the case is to make
use of the light with a "white" intensity spectrum (the light with
its intensity modulated by a "white noise").  Then, at the exit of
the medium, the light intensity spectrum (originally flat) will
show a peak (or a dip) centered at zero frequency with the width ~
$\tau^{-1}$. Usually, however, the frequency dependences of this
kind are obtained using the light with "monochromatic" intensity
spectrum: the light intensity is subjected  to a weak harmonic
modulation, with the phase and amplitude of the modulation being
measured at the exit of the medium as a function of frequency.
Thus, the dependences presented in Fig. 1 can be considered as a
result of filtration of the light intensity spectrum by a
saturable absorber.

\section*{PULSE DELAY AS A RESULT OF GROUP VELOCITY REDUCTION }

Let us now turn to a more sophisticated model that connects the
pulse delay in a saturable absorber with changing group velocity
of light. The effect of coherent population oscillations (CPO)
lying in the basis of this model exploits the same aforementioned
frequency characteristics of the response (Fig. 1a and 1b).  Now,
however, they are implied to demonstrate changes in {\it optical
spectrum} of the light transmitted through the saturable absorber,
rather than changes in its {\it intensity spectrum}.

The CPO effect is observed as follows. A saturable absorber pumped
by a monochromatic light wave (of partially saturating intensity)
with the frequency $\nu_0$ is illuminated by a weak (probe)
monochromatic wave with a shifted frequency of $\nu_0  + \delta
\nu$ (Fig. 2а). As a result of beats of the two waves with close
frequencies, intensity of the total light field appears to be
modulated at the difference frequency $\delta\nu$ (Fig. 2b). We
know, however (see Fig. 1a), that the light intensity modulation
depth at the exit of the saturable absorber changes when the
modulation frequency is comparable with the inverse relaxation
time of the absorber. In particular, the bleachable absorber
amplifies modulation of the transmitted light (Fig. 2c). As a
result, this inevitably affects {\it optical spectrum}  of the
field. There arise two symmetric sidebands $\nu_0 \pm \delta \nu$,
so that one of them exactly coincides with the probe wave both in
phase and in frequency, thus leading to enhancement of the latter
at the exit of the absorber (Fig. 2d). As a result of this
two-wave interaction, the transmission spectrum of the probe wave
in the vicinity of the pump acquires the shape of a peak with the
width  $\sim\tau^{-1}$, centered at the frequency of the pump
$\nu_0$ (Fig. 3a). This is the gist of the CPO effect. The hole in
the absorption spectrum of the medium burnt in this way is, at low
intensities of the probe wave does not depend on the probe beam
intensity, and the saturable absorber in the field of a resonant
monochromatic pump  can be considered as a linear "supermedium"
with a narrow dip in the absorption spectrum at the frequency of
the pump. If this is the case, then such a medium, in conformity
with the known Kramers-Kronig relations, should exhibit, in the
vicinity of the dip, the region of a highly steep dispersion (Fig.
3b).

Recall that at high steepness of dispersion of the refractive
index $n$ ($\omega \frac{dn}{d\omega} \gg 1$), the value of the
light group velocity in the medium

\begin{equation}
V_{gr}= \frac{c}{n + \omega (dn/d\omega)}
\end{equation}
is determined by this {\it dispersion-related} contribution. Thus,
the saturable absorber, in the framework of such a model, allows
one to realize situation similar to that used in the first
successful experiments on slow light based on the effect of
electromagnetically induced transparency. \cite{22, 23}. In this
case, the group velocity of the light pulse, with its spectrum
lying within the width of the dip, should strongly differ from the
phase velocity of light in the medium. In particular, in the
bleachable absorber, dispersion of the refractive index in the
vicinity of the central frequency of the dip will be {\it normal}
($\frac{dn}{d\omega} > 0$), and, correspondingly, the group
velocity of the "resonant" probe pulse will be anomalously low
("slow light"), while in the reverse absorber with anomalous
dispersion in the region of the dip ($\frac{dn}{d\omega} < 0$),
group velocity of the light will be anomalously high ("fast
light").

This model was used to interpret the experiments carried out with
the ruby and alexandrite crystals \cite{8, 9} that were laid into
the basis of the new slow light technique.

It should be noted, however, that this model implies another type
of the experiment. While in the first case (pulse distortion in a
saturable absorber) one studies optical response of the medium to
intensity modulation of the light with arbitrary spectrum, the new
model deals with the response of the medium pumped by a
monochromatic (quasimonochromatic) field to the light pulse with a
sufficiently narrow spectrum, resonant with respect to the pump.
Therefore, these two models, strictly speaking, describe different
experiments. In reality, however, the second model was used to
interpret simple single-beam experiments with saturable absorbers.
This has eventually lead to collision of the two models.

\section*{DISCUSSION}

The start of this collision was given by the papers on "hole
burning" in homogeneously broadened absorption spectra of the ruby
and alexandrite crystals \cite{24, 25}, published long before the
idea of "slow light" was born.

    As was mentioned above, the dip (or the peak) in the modulation
    spectrum of a saturable absorber at zero frequency is its unavoidable
    feature that should be necessarily observed. On the other hand, the effect
    of coherent population oscillations is the manifestation of the same feature,
    but under essentially different experimental conditions, when the intensity
    modulation is formed by superposition  of two fields with close frequencies
    (strong and weak), when one of them (weak) is monitored.  The dip in the intensity
    modulation spectrum by itself, reflecting basic properties of the saturable absorber,
    evidently, cannot serve as an evidence for the dip in the optical spectrum.

Still, the conclusion about spectral hole-burning under conditions
of CPO has been made in \cite{24, 25} based solely on observation
of the dip in the intensity modulation spectrum. According to the
proposed model, the dip in the optical spectrum of the saturable
absorber is detected by sidebands of the modulated laser beam and
is created by the field of carrier frequency. The width of the dip
lied in the range of tens of Hz, and the conclusion made by the
authors evidently could not be confirmed by direct spectroscopic
measurements. In these papers, the questions about spectral width
of the laser source and, therefore, about meeting conditions
needed for observation of the CPO effect, were not discussed.

The idea of using the narrow spectral dip arising under conditions
of CPO for obtaining slow light was borrowed from the papers
\cite{22, 23}, where a similar dip arose under conditions of
electromagnetically induced transparency \cite{8, 9}. The
amplitude measurements of the modulation spectra \cite{24, 25}
were complemented by temporal measurements.  The results of these
measurements completely agreed with the simplest model of the
saturable absorber \cite{2, 4}, but, in accordance with papers
\cite{24, 25}, they were interpreted in terms of the "CPO-based
slow light".

The new interpretation of pulse delay in a saturable absorber had
indeed brought this effect to the front line of science, while the
simplicity of the experiments in combination with the scale of the
easily achievable values "group indices" (corresponding to group
velocities down to fractions of mm/s \cite{26}) rendered this slow
light technique highly popular.

All the research regarding the light pulse delay in saturable
absorbers after that was carried out exclusively in the framework
of the "slow light" concept, which postulated mechanism of the
effect by its title, and as a result, no alternative models of the
pulse delay could be considered.  Meanwhile, the new model did not
look convincing.

The first question arising in the new interpretation of phase
delay of the harmonically modulated light was related to spectral
width of the laser light.  How could the light source with the
line width lying in the range of hundreds MHz create the spectral
dip with a width of several tens of Hz?  Attempts to explain this
paradox were made much later in terms of a special model of the
laser field \cite{16}. It is appropriate to note here that, in
full agreement with regularities of behavior of the saturable
absorber, all the indications of the "CPO-based slow and fast
light" were observed in incoherent light of thermal sources
\cite{27}, when the model of the laser field proposed in \cite{16}
cannot be valid.

The second problem mentioned already in \cite{8} was that the
proposed theory could not explain the delay of a single pulse in
the absence of pump. In authors' opinion, this was a unique case
when the saturating pulse produced its own pump field.

It is noteworthy that practically in all studies of the "CPO-based
slow light" the experiment was performed using a single beam whose
intensity was modulated in a harmonic or pulsed fashion. In other
words, the conditions for observation of the CPO effect (that
imply separation of the probe beam from the pump) were not
satisfied. An exception was the paper \cite{28} that demonstrated
experimentally the fact that the hole burnt by a monochromatic
pump in the spectrum of a saturable absorber under conditions of
CPO is correlated, in conformity with the Kramers-Kronig
relations, with  a dispersion curve, highly steep at the center of
the hole. Direct measurements of the light pulse delay, in this
paper, were not performed. An attempt to observe the CPO-based
group velocity reduction with separation of the pump and probe
beams was made in \cite{29}. However, the object for study (an
ensemble of quantum dots) was characterized by a strong
inhomogeneous broadening, and, for this reason, the experiment was
reduced to detection of spectral hole in the inhomogeneously
broadened absorption band with observation of the related slow
light as in \cite{30}. In recent years, a considerable number of
papers have been published that convincingly showed that all the
known experiments on the "CPO-based slow light" can be described,
in a more natural way, in the framework of the simplest model of
saturable absorber, without invoking the effect of coherent
population oscillations, spectral hole burning, and anomalous
modifications of the light group velocity \cite{7, 13, 14, 17, 18,
20, 31}.  In some papers, there have been described experiments
that came into direct conflict with the "slow light" model. In
particular, in \cite{20}, there has been demonstrated possibility
to control the light pulse delay in a saturable absorber using the
pump beam strongly shifted with respect to that of the probe
light. In \cite{27}, there have been demonstrated all the features
of the "light with negative group velocity" and "ultraslow light"
on simple photochromic objects using an incoherent light source
(incandescent lamp), when the condition for the CPO definitely
could not be met. Universality of the effects of delay of
modulated signals in the nonlinear systems described by Eqs.
(1)--(3) has been shown in \cite{7}  using as an example the
electric circuit with a nonlinear resistor, where the propagation
effects were of no importance, and the carrier frequency could be
absent at all.

All these publications, without precluding fundamental feasibility
of the "CPO-based slow light", provide strong evidence that at
present there is no necessity to invoke the CPO-based model for
interpretation of known experiments on pulse delay in a saturable
absorber.

\section*{CONCLUSIONS}

It should be noted that the discussion about two models of pulse
delay in a saturable absorber could arise only for the reason that
the authors of \cite{24, 25} missed the fact that the dependences
they had found could be comprehensively described by the simplest
model of saturable absorber \cite{2,3,4} and did not contain any
novelty.

One of the authors of the "CPO-based slow light", R.Boyd, in his
last review \cite{21} acknowledges that the observed pulse delay
can be indeed understood in the sense of the simplest model of
saturable absorber of 60's \cite{2},  which still, in author's
opinion, is equivalent to the "CPO-based slow light".

On the basis of the above consideration, we come to conclusion
that all the experimental data on the "CPO-based slow light"
completely agree with the simplest model of saturable absorber,
have nothing to do with changing group velocity of light, and do
not need, for their interpretation, to invoke the effect of
coherent population oscillations.  Therefore, the term "CPO-based
slow light" is, in our opinion, inappropriate in interpretation of
the experiments on pulse delay in saturable absorber.

\section*{ ACKNOWLEDGEMENTS}

The work was supported by the Analytical Departmental
Goal-Oriented Program "Development of Scientific Potential of
Higher School" of 2009 (project no. 2.1.1/1792).

\newpage
\begin{figure}
\begin{center}
\includegraphics [width=10cm]{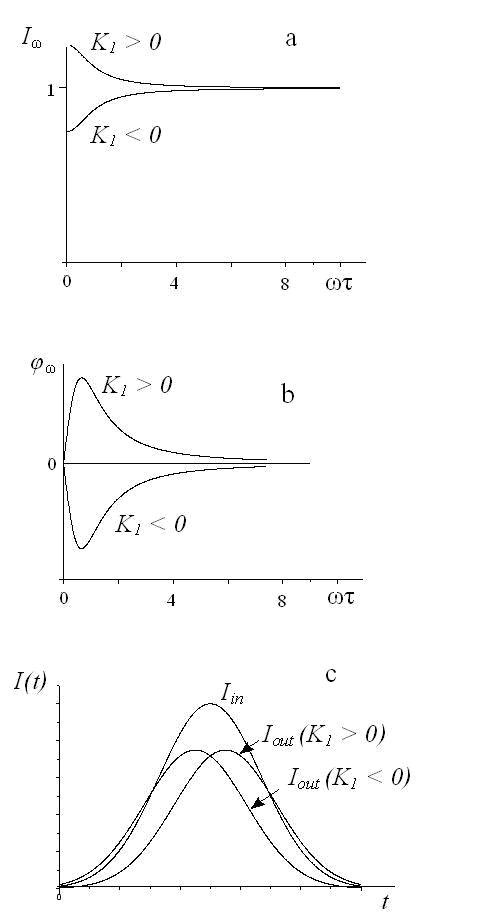}
 \caption{ Frequency dependences of the amplitude ({\it a}) and
phase ({\it b}) of the saturable absorber modulation spectrum and
distortion of the pulse shape ({\it c}).}
 \label{fig1}
 \end{center}
\end{figure}
\begin{figure}
\begin{center}
\includegraphics [width=10cm]{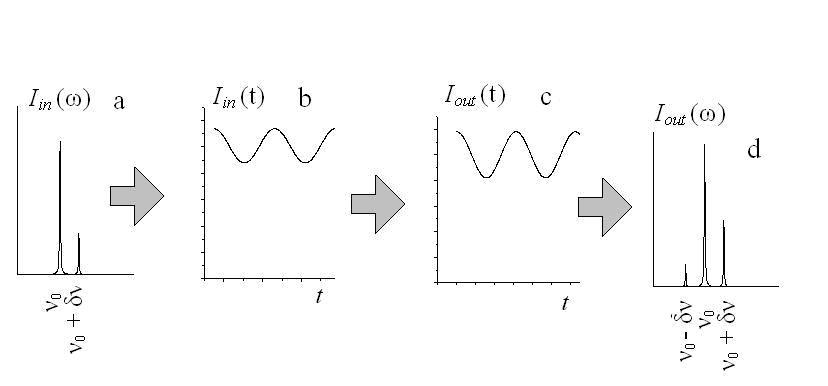}
 \caption{Evolution of optical spectrum of the biharmonic field in
the effect of coherent population oscillations.  {\it a} - optical
spectrum of the original field, {\it b} - time dependence of the
intensity at the entrance of the absorber, {\it c} - time
dependence of the intensity at the exit of the absorber
(increasing modulation depth), {\it d} - optical spectrum of the
field at the exit. }
 \label{fig2}
 \end{center}
\end{figure}
\begin{figure}
\begin{center}
\includegraphics [width=10cm]{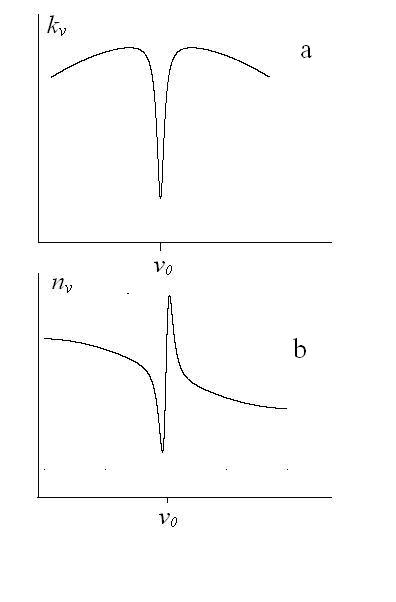}
 \caption{Absorption ({\it a}) and refraction ({\it b}) spectra of a
bleachable absorber detected by the probe light under conditions
of CPO ($\nu_0$ is the pump field frequency). }
 \label{fig3}
 \end{center}
\end{figure}

\end{document}